\documentclass[11pt,journal,epsfig,onecolumn,peerreviewca,draftclsnofoot]{IEEEtran}
\usepackage[dvips]{graphicx}
\usepackage{graphicx}
\usepackage[centerlast]{caption2}
\usepackage{amssymb}
\usepackage{cite}
\usepackage{float}
\usepackage{amsmath}

\begin{document}

\title{Real-Valued Khatri-Rao Subspace Approaches \\on the ULA and a
New Nested Array}

\author{\authorblockN{Huiping~Duan, Tiantian Tuo, Jun Fang and Bing Zeng}
\thanks{Huiping Duan, Tiantian Tuo and Bing Zeng are with the School of Electronic Engineering, University of Electronic Science and Technology of China, Chengdu, China, 611731, Email:
huipingduan@uestc.edu.cn, 201321020471@std.uestc.edu.cn,
eezeng@uest.edu.cn.}
\thanks{Jun Fang is with the National Key Laboratory of Science
and Technology on Communications, University of Electronic Science
and Technology of China, Chengdu, China, 611731, Email:
JunFang@uestc.edu.cn.}
\thanks{This work was supported in part by the National Natural Science Foundation of China under Grants 61201274, 61172114, 61370148.}}

\maketitle

\begin{abstract}
In underdetermined direction-of-arrival (DOA) estimation using the
covariance-based signal models, the computational complexity turns
into a noticeable issue because of the high dimension of the virtual
array manifold. In this paper, real-valued Khatri-Rao (KR)
approaches are developed on the uniform linear array (ULA) and the
nested array. The complexities of subspace decomposition and
spectral search are reduced compared with the complex-valued KR
approach. By designing a special transformation matrix, the
influence of the noise is removed in the mean time while the data is
transformed from the complex domain to the real domain. Deploying
the sensors with nonuniform spacings can raise the degree of freedom
(DOF) and hence help detect more sources in the underdetermined
situation. To increase the DOF further, a new nested array geometry
is designed. The real-valued denoising KR approach developed on the
new nested array can resolve more sources with reduced complexities.
The performance improvement is demonstrated by numerical studies.
\end{abstract}


\section{Introduction} \label{sec1}
In source localization using antenna arrays in radar, sonar and
communication systems, the underdetermined situation
\cite{Van2002,Haykin1992,He2015}, where the number of sources
exceeds the number of sensors, has been paid special attentions.

In order to detect more sources, covariance-based algorithms have
been explored. In these algorithms, the degree of freedom (DOF),
which is measured by the number of distinct cross correlation terms
in the associated difference co-array \cite{Pal2010}, plays an
important role in determining the number of sources the array can
identify. In \cite{Ma2010}, the Khatri-Rao (KR) subspace approach is
developed. The DOF of an N-element uniform linear array (ULA) is
increased to $2N-1$ by exploiting the self-Khatri-Rao product
structure of the array manifold matrix and $2N-2$ sources can be
identified with the KR-MUSIC algorithm. The computational complexity
turns into a noticeable issue because of the high dimension of the
virtual array manifold. Although a dimension reduction strategy is
adopted in \cite{Ma2010}, subspace decomposition and spectral search
are still computationally expensive due to the complex-valued
operations in the algorithm.

Deploying the sensors with nonuniform spacings can raise the DOF and
hence help the covariance-based algorithms detect more sources in
the underdetermined situation. Nonuniform linear arrays, like the
minimum redundancy array (MRA) \cite{Moffet1968} and the
non-redundant array (NRA) \cite{Haykin1992}, have been designed to
enhance the DOF. However, no general analytical formulations can be
provided to express the array geometry or the DOF due to the lack of
regularity in sensor deployment in MRA and NRA. Lately, two types of
nonuniform linear arrays, the co-prime array and the nested array,
are proposed in \cite{Pal2010, Pal2011}. The nested array consists
of two or more ULAs with increasing intersensor spacings. The DOF
achieved by using an N-element nested array is $\mathcal{O}(N^2)$.
Compared with MRA and NRA, the uniform geometry inside each level of
the nested array simplifies the formulation and analysis but
sacrifices the DOF. It is desirable to achieve higher DOF by
inerratic array geometry.

In this paper, real-valued KR approaches are developed for the
underdetermined direction-of-arrival (DOA) estimation problem. The
complexities of subspace decomposition and spectral search are
reduced compared with the complex-valued KR approach. By designing a
special transformation matrix, the influence of the noise is removed
in the mean time while the data is transformed from the complex
domain to the real domain. Unlike using the orthogonal complement
projecting \cite{Ma2010} or eliminating entries of the covariance
matrix by additional matrix multiplication \cite{He2014}, no extra
operations are required for eliminating the noise in the proposed
real-valued KR approaches. By relocating the origin of coordinate
and increasing the sensor spacing of the outer-level ULA of Pal's
nested array, a new nested array geometry is designed. The
real-valued denoising KR approach developed on the new nested array
can resolve more sources with reduced computational complexity and
increased spectral search efficiency.

The rest of the paper is organized as follows: the signal model
based on the KR product is described in Section \ref{Sec2}. In
Section \ref{Sec3}, the real-valued KR approach on the ULA is
designed. The new nested array geometry and the real-valued KR
approach developed on it are presented in Section \ref{Sec4}.
Simulation results are demonstrated in Section \ref{Sec5}. Finally,
the paper is concluded in Section \ref{Sec6}.

\section{Signal Model based on the KR Product} \label{Sec2}
Consider $K$ narrowband far-field signals impinging on an N-element
uniform or non-uniform linear array. The signals are assumed to be
zero-mean quasi-stationary sources \cite{Ma2010} with locally static
second-order statistics. The array received data at $T$ time
snapshots is modeled as
\begin{equation}\label{equ:x}
{\bf{x}}(t) = {\bf{As}}(t) + {\bf{v}}(t), \quad t =0,1, \ldots,T-1
\end{equation}
with ${\bf{x}}(t) = {[{x_1}(t)  \cdots {x_N}(t)]^T}$ and
${\bf{s}}(t) = {[{s_1}(t) \cdots {s_K}(t)]^T}$. $x_n(t)$ is the data
received by the $n$th sensor. $s_k(t), k=1,2,\ldots,K$ are the
sources which are uncorrelated with each other. ${\bf{v}}(t) \in
{\mathbb{C}^{N \times 1 }}$ is the noise assumed to be zero-mean
wide-sense stationary and statistically independent of the source
signals. ${\bf{A}} = [{\bf{a}}({\theta _1}) \phantom{0} \cdots
\phantom{0} {\bf{a}}({\theta _k})] \in {\mathbb{C}^{N \times K}}$ is
the array manifold matrix where ${\theta _k} \in [-\pi /2,\pi /2]$
is the DOA of the $k$th source, and
\begin{equation}
{\bf{a}}(\theta_k ) = {[{e^{ - \frac{{j2\pi {d_1}}}{\lambda }\sin
\theta_k }} \phantom{0} \cdots \phantom{0} {e^{ - \frac{{j2\pi
{d_N}}}{\lambda }\sin \theta_k }}]^T}
\end{equation}
is the $k$th steering vector. Here, $\lambda$ and $d_i$ represent
the signal wavelength and the location of the $i$th sensor,
respectively.

Divide the $T$ snapshots into frames with the frame length $L$. The
local covariance matrix is defined as
\begin{equation}
 {{\bf{R}}_m} = \text{E}\{ {\bf{x}}(t){{\bf{x}}^H}(t)\}, \phantom{0} \forall t \in [(m - 1)L,mL - 1],
 \end{equation}
where $m \in [1,M]$ represents the frame index. ${\bf{R}}_m$ can be
formulated as
\begin{align}
{{\bf{R}}_m} &= {\bf{A}}{{\bf{D}}_m}{{\bf{A}}^H}{\rm{ + }}{\bf{C}} \notag \\
 &= {\bf{A}}\left[ {\begin{array}{*{20}{c}}
{\sigma_{m1}}&{0}&{\cdots}&{0}\\
{0}&{\sigma_{m2}}&{\cdots}&{0}\\
{\vdots}&{\vdots}& \ddots &{\vdots}\\
{0}&{0}&{\cdots}&{\sigma _{mK}}
\end{array}} \right]{{\bf{A}}^H} + {\bf{C}},
\end{align}
where ${\sigma_{mk}} = \text{E} \{ {\left| {{s_k}(t)} \right|^2}\} $
is the source power, and $\bf{C}=E\{ {\bf{v}}(t){{\bf{v}}^H}(t)\}$
is the noise covariance matrix.

Following \cite{Ma2010}, $\textbf{R}_m$ is vectorized:
\begin{align}
{{\bf{y}}_m} &= \text{vec}({{\bf{R}}_m}) = \text{vec}({\bf{A}}{{\bf{D}}_m}{{\bf{A}}^H}) + \text{vec}({\bf{C}}) \notag \\
 &= ({{\bf{A}}^*} \odot {\bf{A}}){{\bf{d}}_m} +
 \text{vec}({\bf{C}}),
\end{align}
where $\odot$ represents the Khatri-Rao product:
\begin{equation}
{{\bf{A}}^*} \odot {\bf{A}} = [{{\bf{a}}^*}({\theta _1}) \otimes
{\bf{a}}({\theta _1}) \phantom{0} \cdots \phantom{0}
{{\bf{a}}^*}({\theta _k}) \otimes {\bf{a}}({\theta _k})]
\end{equation}
with $\bigotimes$ denoting the Kronecker product and
${\bf{d}}_m=\text{Diag}\{{\bf{D}}_m\}$. Letting ${\bf{Y}} =
[{{\bf{y}}_1} \phantom{0} \cdots \phantom{0} {{\bf{y}}_M}]$, we have
\begin{equation} \label{equ:Y}
{\bf{Y}} = ({{\bf{A}}^*} \odot {\bf{A}}){{\bf{\Psi }}^T} +
\text{vec}({\bf{C}}){\bf{1}}_M^T,
\end{equation}
where ${{\bf{1}}_M} = {[1 \phantom{0} \cdots \phantom{0}1]^T} \in
{R^M}$ and
\begin{align}
{\bf{\Psi }} &= {[{{\bf{d}}_1} \phantom{0} \cdots \phantom{0} {{\bf{d}}_M}]^T} \notag \\
 &= \left[ {\begin{array}{*{20}{c}}
{{d_{11}}}&{{d_{12}}}& \cdots &{{d_{1K}}}\\
{{d_{21}}}&{{d_{22}}}& \cdots &{{d_{2K}}}\\
 \vdots & \vdots & \ddots & \vdots \\
{{d_{M1}}}&{{d_{M2}}}& \cdots &{{d_{MK}}}
\end{array}} \right].
\end{align}

Compared with (\ref{equ:x}), $\textbf{Y}$ in (\ref{equ:Y}) is just
like the data received at an array whose manifold matrix is
${{\bf{A}}^*} \odot {\bf{A}}$. Hence, instead of (\ref{equ:x}), DOA
estimation can be carried out based on the Khatr-Rao-product model
(\ref{equ:Y}). According to \cite{Ma2010}, for quasi-stationary
sources with long enough sampling duration and sufficient power
variation, $\text{rank}(\bf{\Psi})=K$ can be satisfied. $\bf{A}^*
\odot \bf{A}$ is full column rank When $K \leq 2N-1$.

\section{Real-Valued KR Approach on the ULA} \label{Sec3}
\indent Consider a linear array with the uniform geometry. Applying
the dimension-reduction idea in \cite{Ma2010}, the virtual array
manifold can be written as
\begin{equation}
{{\bf{A}}^*} \odot {\bf{A}} = {\bf{G\widetilde{B}}},
\end{equation}
where $\textbf{G}={[\textbf{G}_1^T \phantom{0} \textbf{G}_2^T
\phantom{0} \cdots \phantom{0}
\textbf{G}_N^T]^T}\in\mathbb{C}^{N^2\times (2N-1)}$ with all entries
of the matrix $\textbf{G}_i\in \mathbb{C}^{N\times (2N-1)}$ being
zero except
\begin{equation}
\textbf{G}_i(1:N,N+1-i:2N-i)= \textbf{I}_N, i=1,2,\ldots,N,
\end{equation}
where $\textbf{I}_N$ represents the $N \times N$ identity matrix.
${\bf{\widetilde{B}}} = [{\bf{b}}({\theta _1}), \ldots
,{\bf{b}}({\theta _K})] \in {\mathbb{C}^{(2N - 1) \times K}}$ with
\begin{equation}
\begin{array}{ll}
{\bf{b}}(\theta) &= [{e^{(N - 1)\frac{{j2\pi d}}{\lambda }\sin \theta }} \phantom{0} \cdots \phantom{0} {e^{\frac{{j2\pi d}}{\lambda }\sin \theta}} \phantom{0} 1 \cdots \\
&\phantom{=} {e^{ - (N - 1)\frac{{j2\pi d}}{\lambda }\sin \theta}}].
\end{array}
\end{equation}
Let $\textbf{W}=\textbf{G}^T\textbf{G}$. It can be derived that
\begin{equation}
\textbf{W} = \text{Diag}\big\{[1 \phantom{0} 2 \phantom{0} \cdots
\phantom{0} N - 1 \phantom{0} N \phantom{0} N - 1 \phantom{0} \cdots
\phantom{0} 2 \phantom{0} 1]\big\}.
\end{equation}
The dimension of $\textbf{Y}$ in (\ref{equ:Y}) can be reduced by a
linear transformation:
\begin{align} \label{equ:Y2}
{\bf{\tilde Y}} &= {{\bf{W}}^{ - 1/2}}{{\bf{G}}^T}{\bf{Y}} \notag \\
 &= {{\bf{W}}^{1/2}}{\bf{\tilde B}}{{\bf{\Psi }}^T} + {{\bf{W}}^{ - 1/2}}{{\bf{G}}^T}\text{vec}({\bf{C}}){\bf{1}}_M^T.
\end{align}
In fact, by left multiplying $\bf{W}^{- 1/2}{\bf{G}}^T$ on $\bf{Y}$,
the repeated rows of ${{\bf{A}}^*} \odot {\bf{A}}$ are averaged and
sorted. Let $\textbf{B}={{\bf{W}}^{1/2}}{\bf{\tilde B}}$ be the
virtual array manifold after dimension reduction. To separate the
real and the imaginary parts, (\ref{equ:Y2}) is rewritten as
\begin{equation} \label{equ:Y3}
\begin{array}{l}
{{{\bf{\tilde Y}}}_R} + j{{{\bf{\tilde Y}}}_I} = ({{\bf{B}}_R} + j{{\bf{B}}_I}){{\bf{\Psi }}^T} + \\
{{\bf{W}}^{ - 1/2}}{{\bf{G}}^T}\text{vec}({\bf{C}}){\bf{1}}_M^T
\end{array}.
\end{equation}
Define two matrices $\textbf{H}_1 \in \mathbb{R}^{(N-1)\times
(2N-1)} $ and $\textbf{H}_2 \in \mathbb{R}^{(N-1)\times (2N-1)} $ as
follows:
\begin{equation}
{{\bf{H}}_1} = \frac{1}{\sqrt{2}}\left[ {\begin{array}{*{20}{c}}
{\underbrace {\begin{array}{*{20}{c}}
1& \cdots &0\\
 \vdots & \ddots & \vdots \\
0& \cdots &1
\end{array}}_{N - 1}}&{\begin{array}{*{20}{c}}
0\\
 \vdots \\
0
\end{array}}&{\underbrace {\begin{array}{*{20}{c}}
0& \cdots &1\\
 \vdots & {\mathinner{\mkern2mu\raise1pt\hbox{.}\mkern2mu
 \raise4pt\hbox{.}\mkern2mu\raise7pt\hbox{.}\mkern1mu}} & \vdots \\
1& \cdots &0
\end{array}}_{N - 1}}
\end{array}} \right], \notag
\end{equation}
\begin{equation}
{{\bf{H}}_2} = \frac{1}{{\sqrt{2}j}}\left[ {\begin{array}{*{20}{c}}
{\underbrace {\begin{array}{*{20}{c}}
1& \cdots &0\\
 \vdots & \ddots & \vdots \\
0& \cdots &1
\end{array}}_{N - 1}}&{\begin{array}{*{20}{c}}
0\\
 \vdots \\
0
\end{array}}&{\underbrace {\begin{array}{*{20}{c}}
0& \cdots &{ - 1}\\
 \vdots & {\mathinner{\mkern2mu\raise1pt\hbox{.}\mkern2mu
 \raise4pt\hbox{.}\mkern2mu\raise7pt\hbox{.}\mkern1mu}} & \vdots \\
{ - 1}& \cdots &0
\end{array}}_{N - 1}}
\end{array}} \right].
\end{equation}
Multiply $\bf{H}_1$ and $\bf{H}_2$ on the left side of equation
(\ref{equ:Y3}):
\begin{equation} \label{equ:Y4}
\begin{array}{l}
{{\bf{H}}_1}{\bf{\tilde Y}} = {{\bf{H}}_1}{\bf{B}}{{\bf{\Psi }}^T} + {{\bf{H}}_1}{{\bf{W}}^{ - 1/2}}{{\bf{G}}^T}\text{vec}({\bf{C}}){\bf{1}}_M^T,\\
{{\bf{H}}_2}{\bf{\tilde Y}} = {{\bf{H}}_2}{\bf{B}}{{\bf{\Psi }}^T} +
{{\bf{H}}_2}{{\bf{W}}^{ -
1/2}}{{\bf{G}}^T}\text{vec}({\bf{C}}){\bf{1}}_M^T.
\end{array}
\end{equation}
As we can see, $\textbf{H}_1 \textbf{B}=\textbf{B}_R$ and
$\textbf{H}_2 \textbf{B}=\textbf{B}_I$ are the real part and the
imaginary part of $\textbf{B}$ multiplied by a scaling coefficient
$\sqrt{2}$, respectively.

For spatially uniform or nonuniform white noise with covariance
matrix $\bf{C}=\text{Diag} \{[\sigma_{n1}^2 \phantom{0}
\sigma_{n2}^2 \phantom{0} \cdots \phantom{0} \sigma_{nN}^2]\}$, we
can derive that ${{\bf{W}}^{ - 1/2}} {{\bf{G}}^T} \text{vec}
({\bf{C}}) =\frac{1}{\sqrt{N}}\sum\limits_{i=1}^{N}\sigma_{ni}^2
\bf{e}=\sigma_n^2 \bf{e}$, where $\bf{e}$ is an $N^2 \times 1$
vector with the Nth element being one and all other elements being
zero. Since the Nth columns of $\bf{H}_1$ and $\bf{H}_2$ are
all-zero vectors, by simple mathematical operations, we can prove
that
\begin{equation}
\begin{array}{l}
{{\bf{H}}_1}{{\bf{W}}^{ - 1/2}}{{\bf{G}}^T}\text{vec}({\bf{C}}){\bf{1}}_M^T = {\bf{0}},\\
{{\bf{H}}_2}{{\bf{W}}^{ -
1/2}}{{\bf{G}}^T}\text{vec}({\bf{C}}){\bf{1}}_M^T = {\bf{0}}.
\end{array}
\end{equation}
Hence, with the transformation matrices $\textbf{\textbf{H}}_1$ and
$\textbf{\textbf{H}}_2$, the additive noise is eliminated and
(\ref{equ:Y4}) can be rewritten as
\begin{equation}
{\bf{\bar Y}} = \left[ {\begin{array}{*{20}{c}}
{{{\bf{H}}_1}}\\
{{{\bf{H}}_2}}\end{array}}\right]{\bf{\tilde Y}}
 = \bf{H} \bf{\tilde Y}=\left[ {\begin{array}{*{20}{c}}
{{{\bf{B}}_R}}\\
{{{\bf{B}}_I}}
\end{array}} \right]{{\bf{\Psi}}^T}.
\end{equation}

Due to the all-zero column in $\bf{H}$, left multiplication of
$\bf{H} {{\bf{W}}^{ - 1/2}}{{\bf{G}}^T}$ lead to the loss of one
degree of freedom and hence $K \leq 2N-2$ has to be satisfied. We
perform the singular value decomposition (SVD) on $\bf{\bar{Y}}$:
\begin{equation}
\bf{\bar{Y}}={U{\Sigma} V}^H,
\end{equation}
where $\textbf{U}\in\mathbb{C}^{(2N-2)\times (2N-2)}$ and
$\textbf{V}\in\mathbb{C}^{M\times M}$ are the left and right
singular matrices, respectively, and $\bf{\Sigma} \in
\mathbb{R}^{(2N-2) \times M}$ contains the singular values in
descending order. Then the noise subspace is estimated as:
\begin{equation}
{\bf{U}}_n=[{\bf{u}}_{K+1},\ldots,{\bf{u}}_{2N-2}]\in\mathbb{C}^{(2N-2)\times
(2N-2-K)}.
\end{equation}
The dimension of ${\bf{U}}_n$ further limits $K$ to be less than
$2N-2$. Finally the spatial spectrum can be calculated as follows:
\begin{equation}
\text{P}(\theta)= \frac{1}{\big\|{\bf{U}_n ^H {\left[
{\begin{array}{*{20}{c}} {{{\bf{b}}_R}^T}&{{{\bf{b}}_I}^T}
\end{array}} \right]^T}}\big\|_2^2},
\end{equation}
where $\bf{b}_R=\bf{H}_1 \bf{W}^{1/2} \bf{b}(\theta)$ and
$\bf{b}_I=\bf{H}_2 \bf{W}^{1/2} \bf{b}(\theta)$.

\section{Real-Valued KR Approach on the Nested Array} \label{Sec4}
A new nested array geometry is designed before developing the
real-valued Khatri-Rao subspace approach on it.

\subsection{The New Nested Geometry}
Similar to Pal's array \cite{Pal2010}, the proposed nested array
consists of two concatenated ULAs which are called the inner and the
outer. The inner ULA has $N_1$ elements with spacing $d_1$ and the
outer ULA has $N_2$ elements with spacing $d_2$. The Pal's nested
array sets ${d_2} = ({N_1} + 1){d_1}$ while the new geometry sets
${d_2} = {N_1}{d_1}$. Note that the new geometry puts the origin of
coordinate on the first sensor of the inner level. More
specifically, the sensors' locations of the new nested array are
given by ${S_{inner}} = \{ (m - 1){d_1},m = 1,2, \ldots ,{N_1}\}$
and ${S_{outer}} = \{ (n + 1){N_1}{d_1},n = 1,2, \ldots ,{N_2}\}$.
According to the knowledge of the difference set, if $N_2 \geq 2$,
this nested array is equivalent to a filled ULA with $2({N_2} +
1){N_1}+1$ elements whose positions are given by
\begin{equation}
{S_{ca}} = \{ n{d_1},n =  - \tilde{M}, \ldots ,\tilde{M},\tilde{M} =
({N_2} + 1){N_1}\}.
\end{equation}

\begin{figure}
\renewcommand{\captionlabeldelim}{.~}
\captionstyle{center} \centering
\includegraphics[width=9cm,height=4cm]{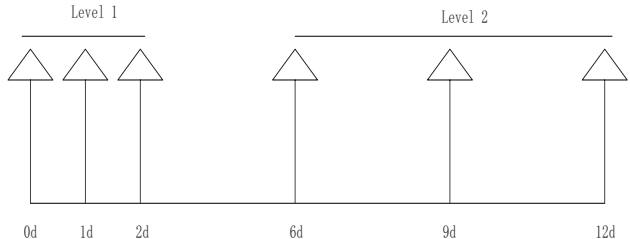}
\caption{\small The proposed two-level nested array with $3$ sensors
in each level.} \label{fig_1}
\end{figure}

Figure \ref{fig_1} shows an example of this new nested array
geometry with $N_1=N_2=3$. As comparisons, in Table \ref{table_one},
the DOF obtained by the MRA \cite{Moffet1968}, the coprime array
\cite{Pal2011}, the ULA (in the KR subspace method \cite{Ma2010}),
the Pal's two-level nested array \cite{Pal2010} and the proposed
two-level nested array are listed for different values of $N_1$ and
$N_2$. The new two-level nested array can attain $2({N_2} +
1){N_1}+1$ DOF using $N_1 + N_2$ elements. When $N_1 \geq N_2$, the
new nested array can increase up to $2(N_1 - N_2)+2$ DOF compared
with Pal's nested array.
\begin{table*}
\newcommand\T{\rule{0pt}{2.6ex}}
\newcommand\B{\rule[-1.2ex]{0pt}{0pt}}
\setlength{\abovecaptionskip}{6pt}
\setlength{\belowcaptionskip}{6pt} \centering \caption{\small
Comparison of the DOF.} \label{table_one} {\small
\begin{tabular}{c|c|c|c|c|c}
\hline $ N_1+N_2$ & MRA & Coprime array & ULA (KR) & Pal's & Proposed \\
\hline 3+2 &19 &6 &9 &15 &19 \\
5+2 &35 &10 &13 &23 &31  \\
5+3 &47 &15 &15 &35 &41 \\
7+3 &73 &21 &19 &47 &57 \\ \hline Formula &inexistent &$N_1\times
N_2$ &$2(N_1+N_2)-1$ &$2N_2(N_1 +1)-1$ &$2(N_2+1)N_1 +1$ \\ \hline
\end{tabular}}
\end{table*}

\subsection{Real-Valued KR Approach on the Nested Array}
\indent Consider model (\ref{equ:Y}) on the new nested array with
$N_1 + N_2$ sensors. The dimension of the virtual array manifold
$\textbf{A}^{*} \odot \textbf{A}$ is $(N_1 + N_2)^{2} \times K$. It
has $2(N_2 +1)N_1 +1$ distinct rows which is as many as the DOF of
the nested array. As what we do for the ULA, the virtual array
manifold $\textbf{A}_1 \in \mathbb{C}^{2(N_2 +1)N_1 +1}$ is
constructed from $\textbf{A}^{*} \odot \textbf{A}$ by averaging the
repeated rows and sorting the elements so that the $i$th row of the
matrix corresponds to the $-(N_2 +1)N_1 +i$th position of the
virtual array. Although it is hard to provide a general expression
for $\bf{G}$ and $\bf{W}$ for the nested array with the arbitrarily
given number of sensors, the operations of averaging and sorting can
be executed on the observation matrix $\textbf{Y}$ in (\ref{equ:Y})
to obtain a new matrix $\textbf{Z}$ as follows
\begin{equation} \label{equ:Z}
{\bf{Z}} = {{\bf{A}}_1}{{\bf{\Psi }}^T} + \sigma _n^2{\bf{\hat e}},
\end{equation}
where $ {\bf{\hat e}} \in {\mathbb{R}^{(2({N_2} + 1){N_1} + 1)
\times M}}$ is a matrix whose elements are all-zero except that the
$(N_2 +1)N_1 +1$th row is an all-one vector.

Two matrices, ${{{\bf{\hat H}}}_1} \in {R^{({N_2} + 1){N_1} \times
(2({N_2} + 1){N_1} + 1)}} $ and ${{{\bf{\hat H}}}_2} \in {R^{({N_2}
+ 1){N_1} \times (2({N_2} + 1){N_1} + 1)}} $, are defined to
transform the complex-valued data into the real-valued one:
\begin{equation}
{{{\bf{\hat H}}}_1} = \frac{1}{\sqrt{2}}\left[
{\begin{array}{*{20}{c}} {\underbrace {\begin{array}{*{20}{c}}
1& \cdots &0\\
 \vdots & \ddots & \vdots \\
0& \cdots &1
\end{array}}_{({N_2} + 1){N_1}}}&{\begin{array}{*{20}{c}}
0\\
 \vdots \\
0
\end{array}}&{\underbrace {\begin{array}{*{20}{c}}
0& \cdots &1\\
 \vdots & {\mathinner{\mkern2mu\raise1pt\hbox{.}\mkern2mu
 \raise4pt\hbox{.}\mkern2mu\raise7pt\hbox{.}\mkern1mu}} & \vdots \\
1& \cdots &0
\end{array}}_{({N_2} + 1){N_1}}}
\end{array}} \right],
\notag
\end{equation}
\begin{equation}
{{{\bf{\hat H}}}_2} = \frac{1}{{\sqrt{2}j}}\left[
{\begin{array}{*{20}{c}} {\underbrace {\begin{array}{*{20}{c}}
1& \cdots &0\\
 \vdots & \ddots & \vdots \\
0& \cdots &1
\end{array}}_{({N_2} + 1){N_1}}}&{\begin{array}{*{20}{c}}
0\\
 \vdots \\
0
\end{array}}&{\underbrace {\begin{array}{*{20}{c}}
0& \cdots &{ - 1}\\
 \vdots & {\mathinner{\mkern2mu\raise1pt\hbox{.}\mkern2mu
 \raise4pt\hbox{.}\mkern2mu\raise7pt\hbox{.}\mkern1mu}} & \vdots \\
{ - 1}& \cdots &0
\end{array}}_{({N_2} + 1){N_1}}}
\end{array}} \right].
\end{equation}
The transformations are as follows:
\begin{equation}
\begin{array}{l}
{{{\bf{\hat H}}}_1}{\bf{Z}} = {{{\bf{\hat H}}}_1}{{\bf{A}}_1}{{\bf{\Psi }}^T} + {{{\bf{\hat H}}}_1}\sigma _n^2{\bf{\hat e}}, \\
{{{\bf{\hat H}}}_2}{\bf{Z}} = {{{\bf{\hat
H}}}_2}{{\bf{A}}_1}{{\bf{\Psi }}^T} + {{{\bf{\hat H}}}_2}\sigma
_n^2{\bf{\hat e}}.
\end{array}
\end{equation}
Since ${{{\bf{\hat H}}}_1} {\bf{\hat e}} = {\bf{0}}$ and
${{{\bf{\hat H}}}_2} {\bf{\hat e}} = {\bf{0}}$, the influence of the
noise is eliminated. Let ${{\bf{A}}_{1R}} = {{{\bf{\hat
H}}}_1}{{\bf{A}}_1}$ and ${{\bf{A}}_{1I}} = {{{\bf{\hat
H}}}_2}{{\bf{A}}_1}$ be the real and the imaginary parts of the
virtual array response matrices. The new model is formulated for the
nested array:
\begin{equation}
{\bf{\bar Z}} = \left[ {\begin{array}{*{20}{c}}
{{{{\bf{\hat H}}}_1}{\bf{Z}}}\\
{{{{\bf{\hat H}}}_2}{\bf{Z}}}
\end{array}} \right] = \left[ {\begin{array}{*{20}{c}}
{{{\bf{A}}_{1R}}}\\
{{{\bf{A}}_{1I}}}
\end{array}} \right]{{\bf{\Psi }}^T}.
\end{equation}
Then SVD can be performed on ${{\bf{\bar Z}}}$ to get the noise
subspace which is applied to search the spectral peaks.

\section{Numerical Studies} \label{Sec5}
Numerical studies are carried out to demonstrate the performance of
the proposed real-valued KR approaches on the ULA and the new nested
array. We consider a ULA with 6 sensors ($N=6$) and a 2-level nested
array with $3$ sensors in each level ($N_1=3,N_2=3$). The
quasi-stationary sources with uniformly distributed random frame
lengths are simulated. The array snapshots are divided into $M$
frames with the frame length $L$ to estimate the local covariance
matrices. The spatial noise is zero-mean and uniformly white complex
Gaussian.

\begin{figure}
\renewcommand{\captionlabeldelim}{.~}
\captionstyle{center} \centering
\includegraphics[width=9cm]{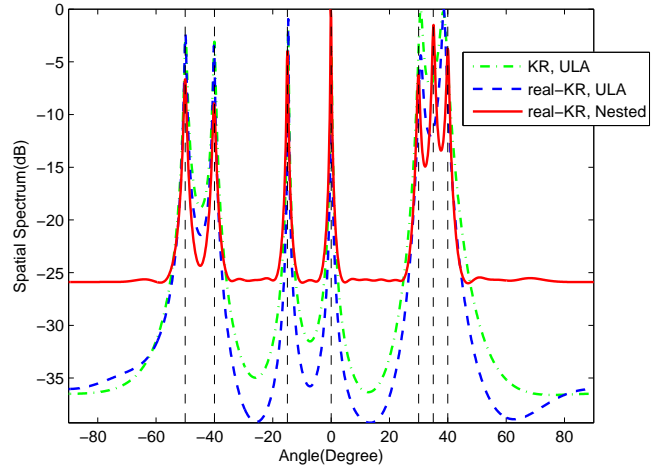}
\caption{\small Spatial spectra in an underdetermined situation.}
\label{fig_2}
\end{figure}

A narrowband underdetermined case with $7$ sources from different
directions, $\{-50^\circ, -40^\circ, -15^\circ, 0^\circ, 30^\circ,\\
35^\circ, 40^\circ\}$, is studied. The signal to noise ratio (SNR)
is set as $0$dB. Simulations use a total of $20000$ snapshots ($T$)
with $400$ snapshots ($L$) in each of the $50$ frame intervals
($M$). The spatial spectra obtained by the KR subspace method and
the real-valued KR approaches on the ULA and the new nested array
are plotted in Figure \ref{fig_2}. We can see that the
complex-valued and the real-valued KR approaches on the ULA provide
similar spatial spectra. The new nested array shows satisfactory
resolving capability while the ULA fails in resolving the closely
spaced sources in this case.

\begin{figure}
\renewcommand{\captionlabeldelim}{.~}
\captionstyle{center} \centering
\includegraphics[width=9.2cm,height=7cm]{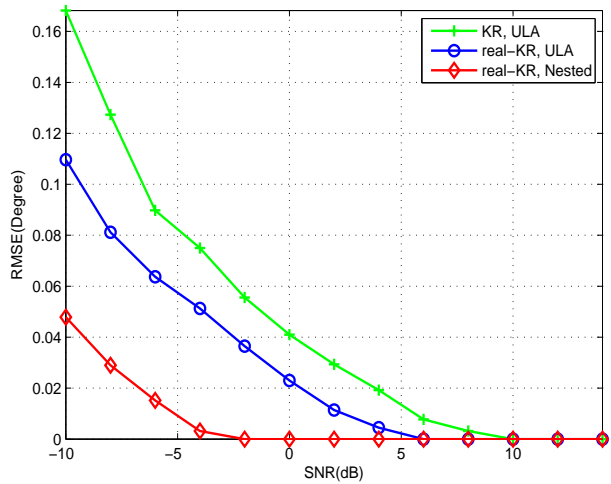}
\caption{\small RMSE vs SNR.} \label{fig_3}
\end{figure}

The root mean square error (RMSE) is evaluated by $1000$ Monte Carlo
simulations. One source from $15^\circ$ is assumed and the SNR
varies from -10dB to 14dB. As shown in Figure \ref{fig_3}, the new
nested array exhibits significantly lower RMSE than the ULA. It is
observed that the real-valued KR subspace method performs better
than the complex-valued KR approach at moderate or low SNR
situations.

In addition, the average time spent in performing SVD and searching
spectral peaks over $100$ trials are listed in Table
\ref{table_two}. The real-valued KR approaches on both the ULA and
the new nested array achieve reduced computational complexity and
increased searching efficiency. This can be explained by the fact
that a complex multiplication is completed by four real
multiplications and two real additions, and a complex addition is
completed by two real additions.
\begin{table}
\newcommand\T{\rule{0pt}{2.6ex}}
\newcommand\B{\rule[-1.2ex]{0pt}{0pt}}
\setlength{\abovecaptionskip}{6pt}
\setlength{\belowcaptionskip}{6pt} \centering \caption{\small SVD
Calculation Time and Spectral Search Time (millisecond).}
\label{table_two} {\small
\begin{tabular}{l|c|c}
\hline Algorithm &SVD calculation time & Spectral search time \\
\hline KR, ULA &0.321 & 6.0\\
real-KR, ULA &0.152 &  3.3\\
KR, Nested &0.678 & 7.4\\
real-KR, Nested &0.621 & 4.7\\
\hline
\end{tabular}}
\end{table}

\section{Conclusions} \label{Sec6}
In this paper, the underdetermined DOA estimation problem is studied
by developing real-valued KR subspace methods and a nonuniform array
geometry. The new nested array geometry can increase the DOF from
$2N_2(N_1 +1)-1$ to $2(N_2 +1)N_1 +1$. Here, $N_1$ and $N_2$ are the
numbers of sensors in the inner and outer levels of the nested
array. The real-valued denoising KR approach developed on the new
nested array can resolve more sources with reduced computational
complexity.


\begin{thebibliography}{1}

\bibitem{Van2002}
H.~L.~V. Trees,
\newblock {\em {D}etection, {E}stimation, and {M}odulation Theory, {P}art {IV},
  {O}ptimum {A}rray {P}rocessing},
\newblock New York: Wiley, 2002.

\bibitem{Haykin1992}
S.~Haykin, J.~P. Reilly, V.~Kezys, and E.~Vertatschitsch,
\newblock ``Some aspects of array signal processing,''
\newblock {\em IEE Proceedings-F}, vol. 139, no. 1, pp. 1--26, Feb. 1992.

\bibitem{He2015}
Z.-Q. He, Z.-P. Shi, L.~Huang, and H.~C. So,
\newblock ``Underdetermined {DOA} estimation for wideband signals using robust
  sparse covariance fitting,''
\newblock {\em IEEE Signal Processing Lett.}, vol. 22, no. 4, pp. 435--439,
  Apr. 2015.

\bibitem{Pal2010}
P.~Pal and P.~P. Vaidyanathan,
\newblock ``Nested arrays: {A} novel approach to array processing with enhanced
  degrees of freedom,''
\newblock {\em IEEE Trans. Antennas Propagat.}, vol. 58, no. 8, pp. 4167--4181,
  Aug. 2010.

\bibitem{Ma2010}
W.-K. Ma, T.-H. Hsieh, and C.-Y. Chi,
\newblock ``{DOA} estimation of quasi-stationary signals with less sensors than
  sources and unknown spatial noise covariance: {A} {K}hatri-{R}ao subspace
  approach,''
\newblock {\em IEEE Trans. Antennas Propagat.}, vol. 58, no. 4, pp. 2168--2180,
  Apr. 2010.

\bibitem{Moffet1968}
T.~A. Moffet,
\newblock ``Minimum redundancy linear arrays,''
\newblock {\em IEEE Trans. Antennas Propagat.}, vol. 16, no. 2, pp. 172--175,
  Mar. 1968.

\bibitem{Pal2011}
P.~Pal and P.~P. Vaidyanathan,
\newblock ``Sparse sensing with co-prime samplers and arrays,''
\newblock {\em IEEE Trans. Signal Processing}, vol. 59, no. 2, pp. 573--586,
  Feb. 2011.

\bibitem{He2014}
Z.-Q. He, Z.-P. Shi, and L.~Huang,
\newblock ``Covariance sparsity-aware {DOA} estimation for nonuniform noise,''
\newblock {\em Digital Signal Processing}, vol. 28, 2014.

\end{thebibliography}
\end{document}